\documentclass[a4paper,fleqn,usenatbib]{mnras}

\usepackage{newtxtext,newtxmath}
\usepackage[T1]{fontenc}
\usepackage{ae,aecompl}

\usepackage{graphicx}	% Including figure files
\usepackage{amsmath}	% Advanced maths commands
\usepackage{amssymb}	% Extra maths symbols

\def\tablefoot#1{\par\vspace*{2ex}%
 \parbox{\hsize}{\leftskip0pt\rightskip0pt
 {\noindent\small\textbf{Notes.}~#1\par}}}
\def\tablefootmark#1{$^{#1}$\,\ignorespaces}
\def\tablefoottext#1#2{$^{(#1)}$~#2}

\usepackage{multirow}
\usepackage{booktabs}

\title[Exploring exomoon atmospheres with a GCM]{Exploring exomoon atmospheres with an idealized general circulation model}

\author[Haqq-Misra and Heller]{Jacob Haqq-Misra$^{1}$\thanks{E-mail: 
jacob@bmsis.org} and Ren\'{e} Heller$^{2}$ \\
$^{1}$Blue Marble Space Institute of Science, 1001 4th Ave Suite 3201, Seattle, WA 98154, USA\\
$^{2}$Max Planck Institute for Solar System Research, Justus-von-Liebig-Weg 3, 37077 G\"ottingen, Germany}

\date{Accepted for publication by \textit{MNRAS} on 18 June 2018}

\pubyear{2018}

\begin{document}
\label{firstpage}
\pagerange{\pageref{firstpage}--\pageref{lastpage}}
\maketitle

\begin{abstract}
Recent studies have shown that large exomoons can form in the accretion disks around super-Jovian extrasolar planets. 
These planets are abundant at about 1\,AU from Sun-like stars, which makes their putative moons interesting for studies of habitability. Technological advances could soon make an exomoon discovery with \textit{Kepler} or the upcoming \textit{CHEOPS} and \textit{PLATO} space missions possible. Exomoon climates might be substantially different from exoplanet climates because the day-night cycles on moons are determined by the moon's synchronous rotation with its host planet. Moreover, planetary illumination at the top of the moon's atmosphere and tidal heating at the moon's surface can be substantial, which can affect the redistribution of energy on exomoons. Using an idealized general circulation model with simplified hydrologic, radiative, and convective processes, we calculate 
surface temperature, wind speed, mean meridional circulation, and energy transport on a 2.5 Mars-mass moon orbiting a 10-Jupiter-mass at 1\,AU from a Sun-like star. The strong thermal irradiation from a young giant planet causes the satellite's polar regions to warm, which remains consistent with the dynamically-driven polar amplification seen in Earth models that lack ice-albedo feedback. Thermal irradiation from young, luminous giant planets onto water-rich exomoons can be strong enough to induce water loss on a planet, which could lead to a runaway greenhouse. Moons that are in synchronous rotation with their host planet and do not experience a runaway greenhouse could experience substantial polar melting induced by the polar amplification of planetary illumination and geothermal heating from tidal effects.
\end{abstract}


%%%%%%%%%%%%%%%%%%%%%%%%%%%%%%%%%%%%%%%%%%%%%%%%%%

%%%%%%%%%%%%%%%%% BODY OF PAPER %%%%%%%%%%%%%%%%%%

\section{INTRODUCTION}
Low-mass stars are conventionally thought to exhibit the most promising odds for the detection of terrestrial planets, at least from an observational point of view. Their low masses enable detections of low-mass companions like Earth-mass planets using radial velocity measurements, and their small radii allow findings of small transiting objects in photometric time series. The recent detections of sub-Earth-sized planets around the M dwarf stars TRAPPIST-1 \citep{Gillon2016,2017Natur.542..456G}, Proxima Centauri \citep{2016Natur.536..437A}, and LHS\,1140 \citep{Dittmann2017} serve as impressive benchmark discoveries.

Even cooler dwarfs exist. Brown dwarfs (BDs), with masses between about 13 and 75 Jupiter masses ($M_{\rm J}$) cannot fuse hydrogen, but their eternal shrinking nevertheless converts significant amounts of gravitational energy into heat for billions of years. From the perspective of BD formation, one can expect that satellites of BDs should commonly form in the dusty, gaseous disks around accreting BDs {\citep{2007MNRAS.381.1597P}}. At the transitional mass regime to giant planets, models of moon formation have shown that Mars-sized moons should commonly form around the most massive super-Jovian gas giant planets \citep{2015ApJ...806..181H,2015A&A...578A..19H}. Photometric accuracies of $10^{-2}$ have now been achieved on BDs using the \textit{Hubble Space Telescope} \citep{2016ApJ...818..176Z}, and an improvement of about one order of magnitude should allow the detection of moons transiting luminous giant planets that can be directly imaged around their host star \citep{2007cabrera,2014ApJ...796L...1H,2016A&A...588A..34H}.

{The search for exomoons has recently become an active area of research that several groups are now competing in, mostly using space-based stellar photometry of exoplanet-exomoons transits \citep{2012ApJ...750..115K,2013A&A...553A..17S,2015PASP..127.1084S,2015ApJ...806...51H} but also using alternatives such as radio emission from giant planets with magnetic fields that are perturbed by moons \citep{2017RAA....17..121L} or space-based coronographic methods such as spectroastrometry \citep{Agol2015}. With the first tentative detection of exomoons or exorings recently claimed in the literature \citep{2012AJ....143...72M,2014ApJ...785..155B,2015ApJ...812...47U,2017AJ....153..193A,2017LPICo2042.4003H,2018AJ....155...36T,2018arXiv180604672R}, a first unambiguous discovery could thus be imminent. Observational biases as well as dynamical constraints will preferentially reveal large, massive moons beyond 0.1\,AU around their star \citep{2006A&A...450..395S,2014ApJ...787...14H}, similar to the moon system that we investigate in this study.
}

%Here we study the main mechanisms affecting the climates on large{, tidally heated} moons around giant, luminous planets in orbit around Sun-like stars.

The first step in estimating the climate conditions on a moon is in the identification of the relevant energy sources. Different from planets, moons receive stellar reflected light from their planetary host \citep{2013AsBio..13...18H} as well as the planet's own thermal emission \citep{2015IJAsB..14..335H}. In particular, a super-Jovian planet's own luminosity can desiccate its initially water-rich moons over several 100\,Myr and make it ultimately uninhabitable. But exomoons around giant planets can also be subject to significant tidal heating \citep{1987AdSpR...7..125R,2006ApJ...648.1196S,2009ApJ...704.1341C,2013AsBio..13...18H}; see Io around Jupiter for a prominent example in the solar system \citep{1979Sci...203..892P}.

For moons around giant planets in the stellar habitable zone (HZ), the reflected plus thermal planetary light onto the moon is a significant source of energy ($\gtrsim10\,{\rm W\,m}^{-2}$) if the moon is closer than about 10 Jupiter radii ($R_{\rm jup}$) to its host planet. Global top-of-atmosphere flux maps showed that eclipses\footnote{With an eclipse we here refer to a moon moving through the stellar shadow cast by the planet.} can decrease the average energy flux on the moon's subplanetary point by tens of ${\rm W\,m}^{-2}$ relative to the moon's antiplanetary hemisphere \citep{2013AsBio..13...18H}. Eclipses can cause a maximum decrease of the globally averaged stellar flux of $\approx6.4\,\%$ at most \citep{2012A&A...545L...8H}.

In analogy to the stellar HZ, \citet{2013AsBio..13...18H} defined a circumplanetary ``habitable edge'' (HE) that is a critical distance to the planet interior to which a moon with an Earth-like atmosphere \citep[mostly N$_2$, small amounts CO$_2$, see][]{1993Icar..101..108K} and a substantial water reservoir experiences a runaway greenhouse effect and is therefore at least temporarily uninhabitable. In top-of-atmosphere energy flux calculations, moons orbiting planets in the stellar HZ encounter an inner HE but no outer HE. Only beyond the stellar HZ does an outer HE appear around the planet \citep{1987AdSpR...7..125R,2014AsBio..14...50H}. One-dimensional latitudinal energy balance models suggest that moons near the outer edge of the stellar HZ can face an outer HE owing to eclipses and an ice-albedo effect if they orbit their star near the outer edge of the HZ \citep{2014MNRAS.441.3513F,2016MNRAS.457.1233F}.

\citet{2013ApJ...774...27H} used the \citet{2013AsBio..13...18H} model to study the effect of global energy flux variations for hypothetical exomoons orbiting four confirmed giant exoplanets in or near the stellar HZ ($\mu$\,Ara\,b, HD\,28185\,b, BD$+$14\,4559\,b, and HD\,73534). Their focus was on the orbital eccentricity of the planet-moon barycenter around the star with the result that fluctuations of tens of ${\rm W\,m}^{-2}$ can occur on moons orbiting on highly eccentric stellar orbits.

Significant improvements in exomoon climate simulations were presented by \citet{2013MNRAS.432.2994F}, who used a 1D latitudinal energy balance model to assess exomoon surface temperatures under the effect of tidal heating and eclipses. \citet{2014MNRAS.441.3513F} advanced this model by also including planetary illumination, and \citet{2016MNRAS.457.1233F} showed yet another update including a global carbonate-silicate cycle and a viscoelastic tidal heating model. In any of the previous studies that estimated surface temperatures on exomoons \citep{2013ApJ...774...27H,2013MNRAS.432.2994F,2014MNRAS.441.3513F,2016MNRAS.457.1233F}, maximum and minimum surface temperatures on exomoons varied by several degrees Kelvin (K) over the circumplanetary orbit at most, while variations due to a moon's changing stellar distance on its circumplanetary orbit or due to eclipses were a mere $\approx0.1$\,K.

General circulation models (GCMs) have been applied to model the effects of eclipses on the atmosphere and surface conditions on Titan, which shows up to 6\,hr long eclipses during $\approx20$ consecutive orbits around Saturn around equinox. \citet{2016P&SS..121...94T} showed that eclipse-induced cooling of Titan's surface, averaged over one orbit around Saturn, is usually $<1$\,K on the pro-Saturnian hemisphere. 

Here we present the first simulations of exomoon climates using an idealized GCM \citep{2015MNRAS.446..428H}. This GCM improves upon previous 1D studies by  allowing us to calculate the energy transfer not only as a function of latitude but also of longitude and height in the exomoon atmosphere. Our main objective is to determine whether exomoon climates are principally different from exoplanet climates, that is: how do planetary illumination and tidal heating contribute to surface habitability in an exomoon atmosphere? {And ultimately, could these climatic effects of the different heat sources possibly be observed with near-future technology?}

\section{Methods}

\subsection{Choice of the simulated systems}

Most of the major moons in the solar system are in synchronous rotation with their host planet; that is, one and the same hemisphere faces the planet permanently (except maybe for libration effects). The star, however, does not a have fixed position in the reference frame of such a moon, as the satellite rotates with its circumplanetary orbital period ($P_{\rm ps}$). Hence, while stellar illumination can be averaged over the day and night side of the moon (longitudinally but not latitudinally), the planet will always shine on the moon's subplanetary point. This is a principal difference between the illumination effects experienced by a planet and a moon.

Furthermore, there will be planet-moon eclipses. But they will only be relevant to the global climate if the moon is in a very close orbit that is nearly coplanar with the circumstellar orbit. Beyond Io's orbit around Jupiter, the orbit-averaged flux decrease will be a few percent at most, and beyond ten planetary radii around a Jovian planet eclipses will be completely negligible \citep{2012A&A...545L...8H}. We therefore neglect planet-moon eclipses in our simulations.

As an additional heat source, we consider geothermal energy, which can be produced via tidal heating, radiogenic decays in the rocky part of the moon, or through release of primordial heat from the moon's accretion. We do not simulate the production of these heat sources in our model, but use interesting and reasonable fiducial values in our GCM simulations.

The moon's mass is chosen as $0.25$ Earth masses ($M_\oplus$), which we consider as an optimal value in terms of exomoon formation around accreting giant planets \citep{2006Natur.441..834C,2014AsBio..14..798H,2015ApJ...806..181H}, exomoon detectability \citep{2009MNRAS.400..398K,2011EPJWC..1101009L,2014ApJ...787...14H,2014ApJ...796L...1H,2015ApJ...813...14K}, and exomoon habitability \citep{2014OLEB...44..239L}.\footnote{However, \citet{2013MNRAS.432.2549A} showed that photometric red noise from stellar variability might make it difficult to detect even Earth-mass moons in the HZ around low-mass stars with \textit{Kepler}.} As these moons should be water-rich, their radii should be about 0.7 Earth radii ($R_\oplus$). 

We consider two moon orbits, one as wide as Europa's orbit around Jupiter ($10\,R_{\rm Jup}$) and one twice that value. 
The former choice is supported by simulations of moon formation around Jupiter-mass planets, which suggest that icy moons can migrate to $\sim$10\,$R_{\rm Jup}$  within the circumplanetary disks \citep{2010ApJ...714.1052S,2012ApJ...753...60O}.
The latter choice is motivated by the recent prediction that the most massive, water-rich moons around super-Jovian planets beyond 1\,AU should form near the circumplanetary ice line at about 20 to $30\,R_{\rm Jup}$ \citep{2015ApJ...806..181H,2015A&A...578A..19H}. The orbital periods related to these two semi-major axes of the satellite orbit ($a_{\rm ps}$) are $1.175$ and $3.324$\,d. We refer to these orbits as our ``short-period'' and ``long-period'' cases, respectively. In all cases, the moon's spin-orbit misalignment with respect to its circumplanetary orbit is assumed to be zero and variations of planetary illumination due to the moon being on an eccentric orbit are neglected.

As for the geothermal heat budget of the satellite ($F_{\rm g}$), we consider three cases of $0$, $10$, and $100\,{\rm W\,m}^{-2}$. In those cases where tidal heating is a major contribution, the highest values will only be reached in very close-in orbits within $\lesssim10\,R_{\rm Jup}$ around our test planet. If the moon under consideration is the only major satellite, tidal processes will usually act to circularize its orbit \citep{1963MNRAS.126..257G}, to erode its obliquity \citep{2011A&A...528A..27H}, and to lock its rotation rate with the orbital mean motion \citep{2016MNRAS.456..665M}. These processes generate tidal heating in the moon, which will gradually decay over time. Tidal surface heating rates on moons could be $>100\,{\rm W\,m}^{-2}$ for about $10^6$\,yr for a single moon on an initially eccentric orbit \citet{2013AsBio..13...18H}. {In this sense, the physical conditions that we model could preferably correspond to young systems rather than evolved systems. The system age, however, is not an input parameter in our models and our results are not restricted to young systems to begin with. In fact, large extrasolar moons have not been detected unambiguously so far, and so it remains unclear whether they are only subject to significant tidal heating when they are young.} If the moon is member of a multi-satellite system, {for example,} then mutual interaction and resonances could {maintain significant orbital eccentricities and} extend this timescale by {{orders of magnitude \citep{2014AsBio..14..798H,2017MNRAS.472....8Z}}.

As for the planetary illumination absorbed by the moon, planet evolution tracks show that the luminosities of young super-Jovian planets 10 times the mass of Jupiter can be as high as $10^{-5}$ to $10^{-3}$ solar luminosities ($L_\odot$), depending on the planet's core mass, amongst others \citep{2013A&A...558A.113M}. Even at the lower end of this range, a moon in a Europa-wide orbit at about $10\,R_{\rm Jup}$ would absorb about $500\,{\rm W\,m}^{-2}$ (maybe for some ten Myr), thereby easily triggering a runaway greenhouse effect on the moon.\footnote{This case is particularly interesting in view of the possible detection of moons around young, self-luminous giant planets via direct imaging with the \textit{European Extremely Large Telescope} \citep{2014ApJ...796L...1H}.} Hence, we consider four cases of planetary illumination onto the satellite's subplanetary point, namely, $0$, $10$, $100$, and $500\,{\rm W\,m}^{-2}$ (the latter one only in the short-period moon orbit). All these cases are summarized in Table~\ref{tab:models}.

\begin{table*}
\caption{\label{tab:models}Simulated exomoon systems: initialization parameters and global average quantities}
\centering
\begin{tabular}{lcccc@{\hskip 0.5in}cccc}
\hline\hline
\multirow{1}{*}{} & \multicolumn{4}{c|}{Initialization parameters} & \multicolumn{4}{c|}{Average quantities} \\
Case & $a_{\rm ps}$ & $P_{\rm ps}$ & $F_{\rm g}$ & $F_{\rm t}$ & $T_{\rm surf}$ & ${\Delta}T_{\rm pole}$ & $F_{\rm OLR}$ & $q_{\rm strat}$ \\
\hline
1\tablefootmark{a} & $20\,R_{\rm Jup}$ & $3.324$\,d & $0\,{\rm W\,m}^{-2}$ & $0\,{\rm W\,m}^{-2}$ & $289.5\,{\rm K}$ & $0.0\,{\rm K}$ & $236.4\,{\rm W\,m}^{-2}$ & $7.6\times10^{-6}$  \\
2 & $20\,R_{\rm Jup}$ & $3.324$\,d & $0\,{\rm W\,m}^{-2}$ & $10\,{\rm W\,m}^{-2}$ & $290.0\,{\rm K}$ & $0.7\,{\rm K}$ & $239.4\,{\rm W\,m}^{-2}$ & $7.6\times10^{-6}$ \\
3 & $20\,R_{\rm Jup}$ & $3.324$\,d & $10\,{\rm W\,m}^{-2}$ & $0\,{\rm W\,m}^{-2}$ & $291.7\,{\rm K}$ & $4.0\,{\rm K}$ & $246.0\,{\rm W\,m}^{-2}$ & $8.3\times10^{-6}$ \\
4 & $20\,R_{\rm Jup}$ & $3.324$\,d & $10\,{\rm W\,m}^{-2}$ & $10\,{\rm W\,m}^{-2}$ & $292.2\,{\rm K}$ & $5.0\,{\rm K}$ & $249.0\,{\rm W\,m}^{-2}$ & $8.1\times10^{-6}$ \\
5 & $20\,R_{\rm Jup}$ & $3.324$\,d & $10\,{\rm W\,m}^{-2}$ & $100\,{\rm W\,m}^{-2}$ & $295.9\,{\rm K}$ & $11.9\,{\rm K}$ & $276.3\,{\rm W\,m}^{-2}$ & $3.9\times10^{-5}$ \\
\hline
6\tablefootmark{b} & $10\,R_{\rm Jup}$ & $1.175$\,d & $0\,{\rm W\,m}^{-2}$ & $0\,{\rm W\,m}^{-2}$ & $287.9\,{\rm K}$ & $0.0\,{\rm K}$ & $234.9\,{\rm W\,m}^{-2}$ & $1.4\times10^{-5}$ \\
7 & $10\,R_{\rm Jup}$ & $1.175$\,d & $0\,{\rm W\,m}^{-2}$ & $10\,{\rm W\,m}^{-2}$ & $288.4\,{\rm K}$ & $0.6\,{\rm K}$ & $237.9\,{\rm W\,m}^{-2}$ & $1.7\times10^{-5}$ \\
8 & $10\,R_{\rm Jup}$ & $1.175$\,d & $10\,{\rm W\,m}^{-2}$ & $0\,{\rm W\,m}^{-2}$ & $290.0\,{\rm K}$ & $3.8\,{\rm K}$ & $244.4\,{\rm W\,m}^{-2}$ & $2.0\times10^{-5}$ \\
9 & $10\,R_{\rm Jup}$ & $1.175$\,d & $10\,{\rm W\,m}^{-2}$ & $10\,{\rm W\,m}^{-2}$ & $290.5\,{\rm K}$ & $4.5\,{\rm K}$ & $247.5\,{\rm W\,m}^{-2}$ & $2.3\times10^{-5}$ \\
10 & $10\,R_{\rm Jup}$ & $1.175$\,d & $10\,{\rm W\,m}^{-2}$ & $100\,{\rm W\,m}^{-2}$ & $294.9\,{\rm K}$ & $13.8\,{\rm K}$ & $275.5\,{\rm W\,m}^{-2}$ & $6.8\times10^{-5}$ \\
11 & $10\,R_{\rm Jup}$ & $1.175$\,d & $10\,{\rm W\,m}^{-2}$ & $500\,{\rm W\,m}^{-2}$ & $310.0\,{\rm K}$ & $45.5\,{\rm K}$ & $398.3\,{\rm W\,m}^{-2}$ & $3.3\times10^{-3}$ \\
12 & $10\,R_{\rm Jup}$ & $1.175$\,d & $100\,{\rm W\,m}^{-2}$ & $500\,{\rm W\,m}^{-2}$ & $321.9\,{\rm K}$ & $60.8\,{\rm K}$ & $482.2\,{\rm W\,m}^{-2}$ & $1.4\times10^{-2}$ \\
\hline
\end{tabular}
\tablefoot{
In all cases the planet-moon binary orbits at 1\,AU from a Sun-like star, $M_{\rm p}=10\,M_{\rm Jup}$, $M_{\rm s}=0.25\,M_\oplus$, and $R_{\rm s}=0.7\,M_\oplus$. 
The parameter $F_{\rm g}$ is {uniform} geothermal heating {at all latitudes.}
{The parameter} $F_{\rm t}$ is absorbed planetary illumination, {with a latitudinal distribution of
$F_{\rm t}\left|\cos\lambda\right|$ when $90^\circ < \lambda < 270^\circ$ and zero otherwise.}
Mean values from the set of simulations are shown for global surface temperature $T_{\rm surf}$, 
change in polar surface temperature ${\Delta}T_{\rm pole}$,
global outgoing longwave radiative flux at the top of the atmosphere $F_{\rm OLR}$, 
and global specific humidity at the 50 hPa level $q_{\rm strat}$.\\
\tablefoottext{a,b}{We also refer to 1 and 4 as our ``slow rotator control'' and ``rapid rotator control'' cases, respectively.}
}
\end{table*}

\subsection{Climate model}
\label{sub:atmosphere}

We use an atmospheric GCM to simulate the climate of an Earth-like moon in orbit around a super-Jovian planet. 
This GCM was developed by the Geophysical Fluid Dynamics Laboratory (GFDL), based upon their `Flexible Modeling System' (FMS), with 
idealized physical components \citep{Frierson2006,Frierson2007,Haqq-Misra2011,2015MNRAS.446..428H}. The dynamical core uses a spectral decomposition method with T42 resolution to solve the Navier-Stokes 
(or `primitive') equations of motion. 
We use a shallow penetrative adjustment scheme to perform convective adjustment in the model \citep{Frierson2007}, 
which provides a computationally efficient method for restoring vertical balance in lieu of a more explicit representation of convective processes. 
The GCM surface is bounded with a diffusive boundary layer scheme and a 50 m thick thermodynamic ocean layer with a fixed heat capacity. This is 
analogous to assuming that the moons are fully covered with a static, uniform ocean\footnote{This is motivated by our assumption that our test 
moons formed in the icy parts of the accretion disks around their giant host planets at several AU from their Sun-like star. The initial H$_2$O 
ice content of the moons would then have liquefied as the planets and moons migrated to about 1\,AU, where $\sim100$ super-Jovian exoplanets are known today.}; hence, topography is neglected.
Our assumption of aquamoon conditions with no topography or ice also means that we neglect ice-albedo feedback. 
These simplifications allow for computational efficiency, and they allow us to examine fundamental changes in climate structure without any positive 
feedback processes causing the model to become numerically unstable.

The GCM includes two-stream gray radiative transfer, which uses a gray-gas radiative absorber with a specified vertical profile to mimic a 
greenhouse effect \citep{Frierson2006}. The model atmosphere is transparent to incoming stellar (\textit{i.e.,} shortwave) radiation, so that incoming starlight penetrates the atmosphere
and warms the surface (with a fixed surface albedo of 0.31 for shortwave radiation). Stellar radiation is averaged across the surface (so there is no diurnal cycle).
Infrared (\textit{i.e.,} longwave) radiation is absorbed by the gray-gas atmosphere in proportion to the optical depth at each model layer, with 
the surface values of optical depth tuned to reproduce an Earth-like value when the GCM 
is configured with Earth-like parameters. 
Furthermore, water vapor is decoupled from the gray radiative transfer scheme, so that water vapor 
feedback is neglected. 
The GCM is also cloud-free, although we remove excess moisture and energy from the atmosphere through large-scale 
condensation to the surface. 

{Even with these idealized assumptions, this GCM remains capable of representing surface and tropospheric temperature profiles, as well as 
the large-scale circulation features, of Earth today \citep{Frierson2006,Frierson2007,Haqq-Misra2011}. The model maintains symmetry
about the equator due to the lack of a seasonal cycle, which can be interpreted as a mean annual climate state. 
The model was originally developed to explore the role of moisture on tropospheric static stability \citep{Frierson2006} and the transport of static energy in moist
climates \citep{Frierson2007}. 
We also note that this model has
been used to demonstrate that a realistic tropospheric profile can be maintained by eddy fluxes alone, even in the absence of stratospheric 
ozone warming \citep{Haqq-Misra2011}.}
{Even so, our application of this model to exomoons} implies that our results should be interpreted qualitatively, as a conservative estimate 
with regard to surface temperature values, runaway greenhouse thresholds, {and large-scale dynamics.}

The use of a gray-gas absorber allows us to avoid the problem of choosing a specific atmospheric composition, 
as any particular choice of greenhouse gases (such as carbon dioxide or methane) will yield a unique GCM solution.
Although many GCM studies of exoplanet atmospheres use band-dependent (\textit{e.g.,} non-gray) radiation with cloud 
parameterizations \citep{2014ApJ...787L...2Y,Kopparapu2016,wolf2017constraints,leconte2013b,popp2016transition,godolt20153d,Haqq-Misra2018}, 
such an approach introduces a new set of free parameters for determining the appropriate mix of greenhouse and inert gases. 
We instead choose to use an idealized gray-gas GCM for our study of exomoon habitability, as others have done for qualitatively exploring the 
runaway greenhouse threshold \citep{ishiwatari2002numerical}. 
Likewise, the assumption that water vapor is radiatively neutral allows our model to remain 
stable at high stellar flux levels without initiating a runaway greenhouse state;
thus, our assumption of a cloud-free atmosphere and a simplified convection scheme limits the use of our GCM for quantitatively determining the runaway greenhouse threshold.
For example, clouds beneath the subplanetary point 
could help to delay the onset of a runway greenhouse \citep{2014ApJ...787L...2Y}, although rapid rotation may weaken this effect \citep{Kopparapu2016}. 
We emphasize that our radiation limits, particular for identifying a runaway greenhouse, should be interpreted 
in a qualitative sense in order to guide more sophisticated investigations with less-idealized GCMs. 

We fix the moon into synchronous rotation with its planet so that the subplanetary point is centered at the moon's equator, and we set
the moon to have a circular orbit and an obliquity of zero. 
We include an additional source of infrared heating at the top 
of the atmosphere, centered on the subplanetary point, which we use to represent heating by the super-Jovian planet in our simulations. 
Specifically, we set the downward infrared flux at the top of the atmosphere equal to $F_{\rm t}\left|\cos\lambda\right|$ when $90^\circ < \lambda < 270^\circ$, 
and zero otherwise (where $\lambda$ is longitude). Because the atmosphere is absorbing to infared radiation (both in upward and downward directions),
this planetary infrared flux is absorbed by the uppermost atmospheric layers, with none of this infrared radiation penetrating
through to the surface in any of our simulations. This upper atmosphere absorption of infrared radiation from the host planet is 
the primary feature that distinguishes an exomoon climate from an Earth-like exoplanet climate.

We also add geothermal heating uniformly at the bottom of the atmosphere, as a representation of tidal heating. The infrared flux
from the surface follows the Stefan-Boltzman law, with the geothermal heating term, $F_{\rm g}$, added as a secondary source of surface warming.
Geothermal heating due to tidal heating may also appear on synchronously rotating exoplanets \citep{2015MNRAS.446..428H}, particularly those with 
highly eccentric orbits. Geothermal heating can contribute to habitability by increasing surface temperature, while it can also 
alter atmospheric circulation patterns by driving stronger poleward transport. Geothermal heating provides a secondary mechanisms
that may contribut to climatic features on exomoons. 

Each case in Table~\ref{tab:models} was initialized from a state of rest and run for a period of $3,000$\,d in total. The first $1,000$\,d were discarded, and the average of the subsequent $2,000$\,d 
of runtime were used to analyze our cases. The model reaches a statistically steady state within about 500\,d of initialization, which takes approximately 
4\,h to complete using a Linux workstation (6 cores at 2.0 GHz). 
In our presentation of results, we refer to our cases with a 1.175\,d rotation rate 
as `short-peroid' and our cases with a 3.324\,d rotation rate as `long-period.' We also refer to our two cases with geothermal heating and absorbed planetary illumination 
set to zero as our `control' cases. Our set of twelve cases provide an overview of the dependence of an exomoon's climate on the properties of its host planet.

\section{Results}

For all of our control and experimental cases, we calculate global average values of surface temperature, outgoing longwave flux at the top of the atmosphere, and 
stratospheric specific humidity (Table~\ref{tab:models}). Our control cases (1 and 6) both show global average surface temperatures similar to that of Earth today. Our experimental 
cases (2-5 and 7-12) show an increase in temperature as geothermal and planetary fluxes increase, with a corresponding increase in stratospheric water vapor and outgoing 
infrared radiation. 

\subsection{Surface habitability}

Surface temperature and winds are shown for the two control cases in Figure~\ref{fig:surftemp_control}, 
with the slow rotator on the top row, the rapid rotator on the middle row, and the difference between the two on the bottom row. 
The change in rotation rate from the slow to rapid rotator results in an increase in the strength of the easterly equatorial jet, which also 
corresponds to an increase in the equator-to-pole temperature contrast. This expected behavior corroborates the classic results of 
\citet{Williams1982}, who find that even slower rotation rates will result in a global meridional circulation cell that spans the entire hemisphere.
Lacking geothermal or planetary heating, these control cases represent Earth-like climate states for a smaller planet at different rotation rates.

%**********************************************
\begin{figure}
\centering
\includegraphics[width=1.\linewidth]{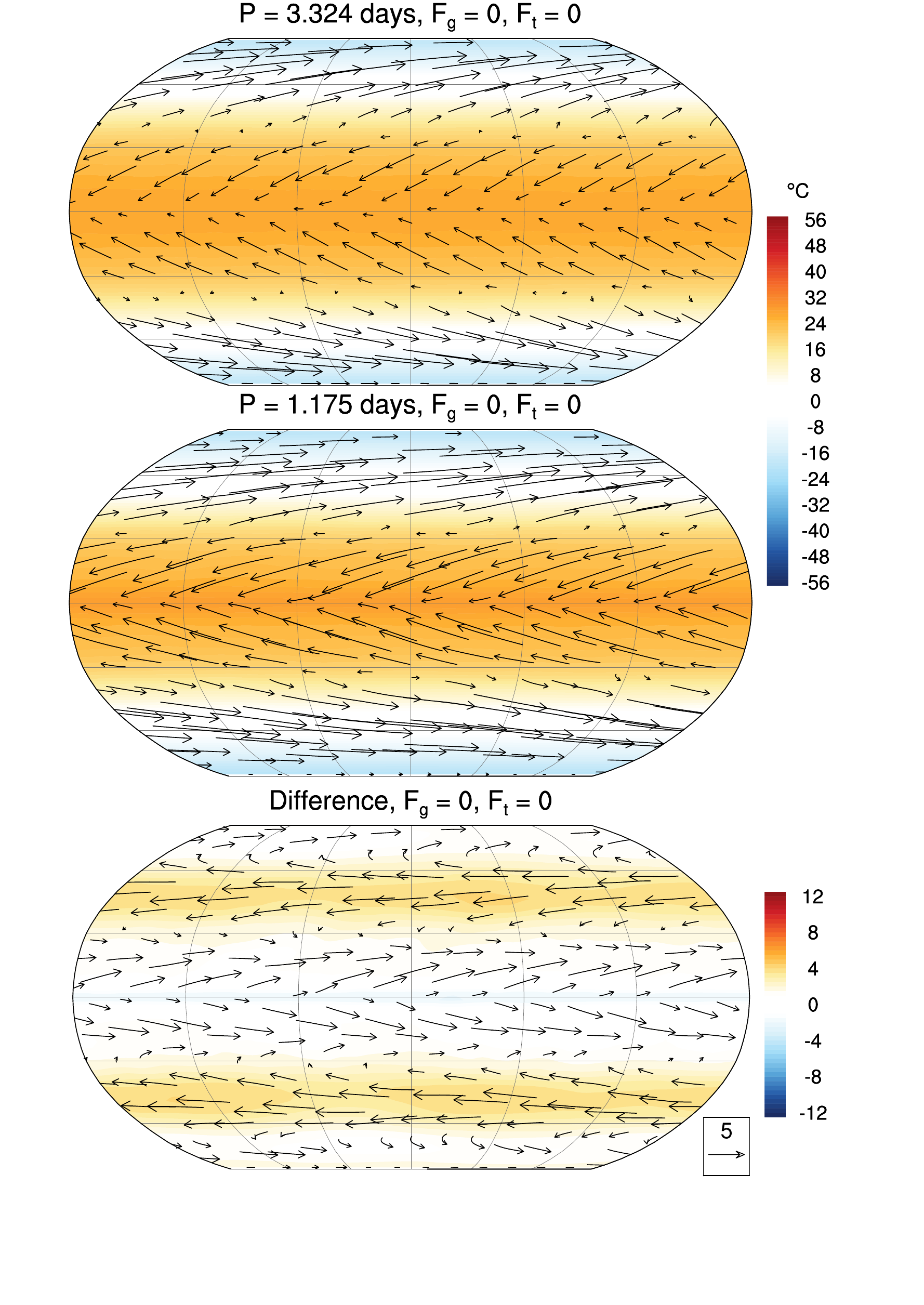}
\caption{Time average of surface temperature deviation from the freezing point of water (shading) and horizontal wind (vectors) for the $P=3.324$\,d (top panel) and $P=1.175$\,d (middle panel) control cases. The bottom row shows the difference of the first row minus the second. The subplanetary point falls on the center of each panel. The lengths of the vectors are proportional to the local wind speeds, and a reference vector with a length of 5\,m\,s$^{-1}$ is shown on the last panel. (The color scale is chosen to match the maximum temperature range shown in Figures \ref{fig:surftemp_short} and \ref{fig:surftemp_longshort} to ease comparison.)}
\label{fig:surftemp_control}
\end{figure}
%**********************************************

%**********************************************
\begin{figure*}
\centering
\includegraphics[angle=90,width=1.\linewidth]{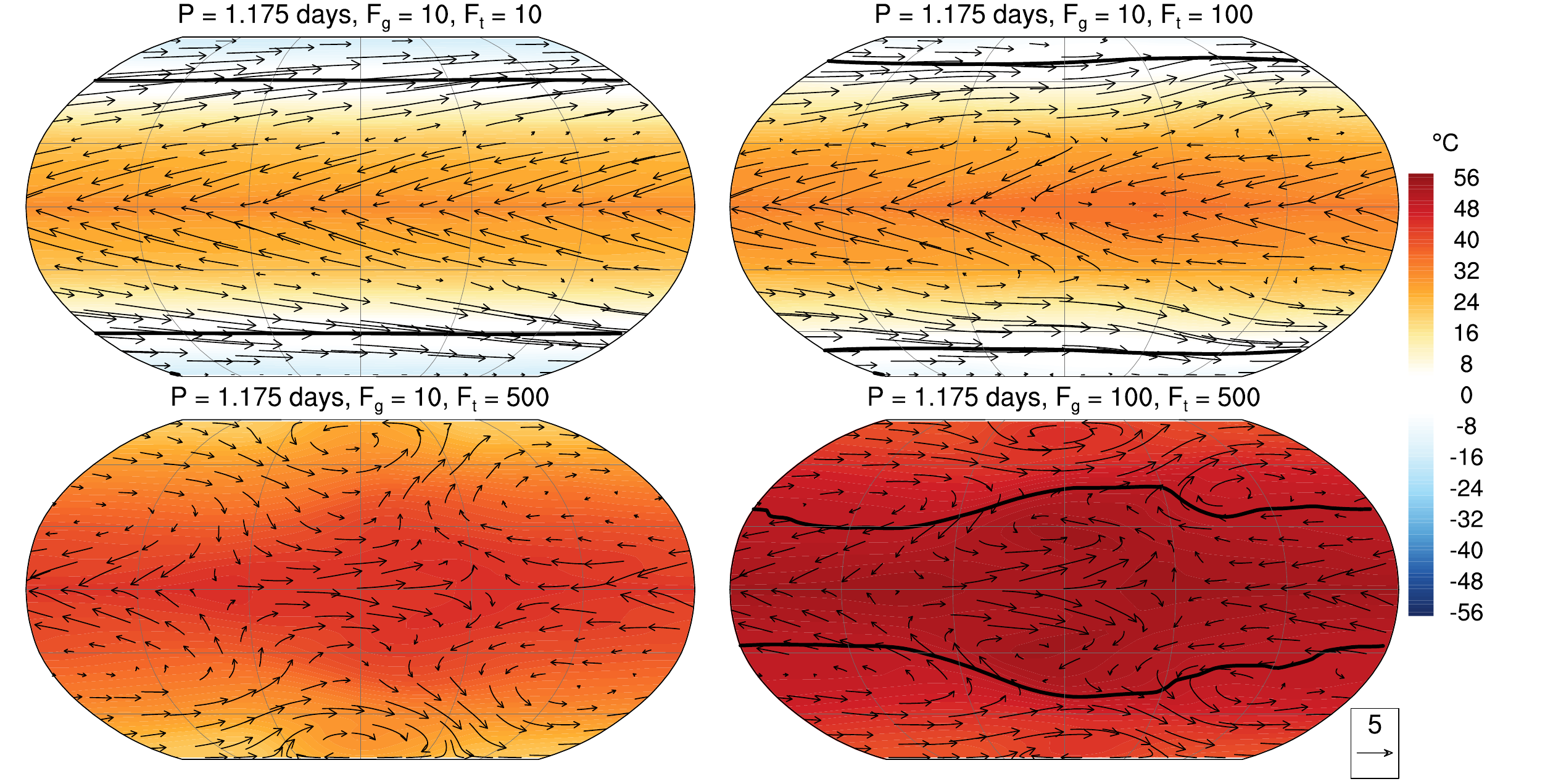}
\caption{Time average of surface temperature deviation from the 
freezing point of water (shading) and horizontal surface wind (vectors) 
for the $P=1.175$\,d experiments. The geothermal ($F_{\rm g}$) and top-of-the 
atmosphere heat flux from the planet ($F_{\rm t}$) are shown on top of each panel. 
Dark coutours indicate the `animal habitability' temperature bounds of 0 and 50 degrees Celsius where complex life could survive.
The lengths of the vectors are proportional to the local wind speeds, and a reference vector with a length of 5\,m\,s$^{-1}$ is shown on the last panel.}
\label{fig:surftemp_short}
\end{figure*}
%**********************************************

The addition of infrared planetary illumination to the top of the atmosphere and geothermal heating to the surface can lead to departures from an
Earth-like climate state. Figure~\ref{fig:surftemp_short} shows surface temperature and winds for the rapid rotator experiments with increasing 
contributions of infrared heating. The top row shows cases with modest geothermal and planetary heating, which leads to both warming of the poles 
and destabilization of the easterly equatorial jet. As additional heating is applied, shown in the bottom rows of Figure~\ref{fig:surftemp_short}, the predominantly 
zonal flow of winds becomes disturbed in favor of a pattern dominated by large-scale vortices. 
Figure~\ref{fig:surftemp_short} also includes dark contours showing the `animal habitable zone' limits of 0 and 50 degrees Celsius, which represents the 
temperature range where complex animal life could survive \citep{ward2000rare,edson2012carbonate}. The two cases in the top row of Figure~\ref{fig:surftemp_short} both show that the freezing 
line of 0 degrees Celsius sits near the boundary between midlatitude and polar regions, around 60 degrees latitude, so that 
the planet's habitable real estate is confined to the midlatitude and equatorial regions. The bottom left panel of Figure~\ref{fig:surftemp_short}
shows a case where the entire surface is within the animal habitability limits, as a result from strong thermal heating from the host planet. The most extreme case,
shown in the bottom right panel of Figure~\ref{fig:surftemp_short}, has the 50 degree Celsius contour at about 30 degrees latitude, with regions closer to the equator
being too warm to support complex life. 

Our set of GCM cases also illustrate the propensity for an exomoon atmosphere to enter a runaway greenhouse as a result of strong thermal heating aloft.
As we discussed above, our idealized GCM neglects water vapor feedback and relies upon gray-gas radiative transfer, 
so our consideration of greenhouse states should be interpreted as a qualitative and conservative estimate. Further GCM development with band-dependent radiative transfer, 
cloud paramterization, and more realistic convective processes will be required to identify quantitative thresholds for when we should 
expect to observe a runaway greenhouse state on an exomoon. 
Additionally, a dry moon with a desert surface and little standing water 
will remain stable past these radiation limits \citep{Abe2011,leconte2013b}, so our moist GCM calculations also serve as a conservative estimate of the runaway greenhouse threshold.
Given these caveats, it still remains instructive to consider how our exomoon simulations compare with theoretical limits for expecting a runaway greenhouse.

The long-period cases (1-5) show a maximum outgoing infrared flux of 276.3\,W\,m$^{-2}$, which falls within the stable radiation limits 
calculated from one-dimensional \citep{Kasting1988,Goldblatt2012,Ramirez2014} and three-dimensional \citep{leconte2013,wolf2014} climate models.
By contrast, the short-period cases (6-12) include experiments with an outgoing infrared flux of 400\,W\,m$^{2}$ or larger (cases 11 and 12), which exceeds stable 
radiation limits and indicates the climate should be in a runaway greenhouse state \citep{Kasting1988,Goldblatt2012,Ramirez2014,leconte2013}. 
The two cases (11 and 12) with $F_{\rm t} = 500$\,W\,m$^{-2}$ are also both within the runaway greenhouse regime, which suggests that their surface
temperatures, as shown in the bottom row of Figure~\ref{fig:surftemp_short}, would continue to increase in a GCM with raditively-coupled water vapor.

Water loss can also occur prior to the runaway greenhouse, due to the photodissociation of water vapor as the stratosphere becomes wet, 
in a process sometimes known as a `moist greenhouse' \citep{Kasting1988,0004-637X-845-1-5}. The moist greenhouse state can be inferred
by amount of water vapor that crosses the tropopause and reaches the stratosphere, which we indicate using specific humidity (the ratio of 
moist to total air), $q_{\rm strat}$, near the model top (see Table~\ref{tab:models}). Calculations with other models indicate that 
atmospheres enter a moist greenhouse and begin to rapidly lose water to space when stratospheric specific humidity exceeds 
a threshold of $q_{\rm strat} \approx10^{-3}$ \citep{Kasting1988,kopparapu2013,wolf2014,Kopparapu2016}. 
As a qualitative approach to this problem, our results illustrate
that thermal heating from the host planet, as well as geothermal heating from tides, are both plausible mechanisms for heating an exomoon 
atmosphere to the point of initiating water loss.

\subsection{Polar amplification of warming}

We compare our potentially habitable long-peroid and short-peroid cases with $F_{\rm g} = 10$\,W\,m$^{-2}$ and $F_{\rm t} = 100$\,W\,m$^{-2}$ with the corresponding control 
cases in Figure~\ref{fig:surftemp_longshort}. These cases are within stable radiation limits and are not at risk of entering a runaway greenhouse or otherwise 
losing water to space due to a wet stratosphere. Both panels in Figure~\ref{fig:surftemp_longshort} show that warming is concentrated toward the poles, 
with a more modest degree of warming at the tropics and midlatitudes. This `polar amplification' is a well-known process that occurs in climate models, 
particularly in simulations of global warming, and is an expected consequence of imposing additional heating on an atmosphere.

%**********************************************
\begin{figure*}
\centering
\includegraphics[angle=90,width=1.\linewidth]{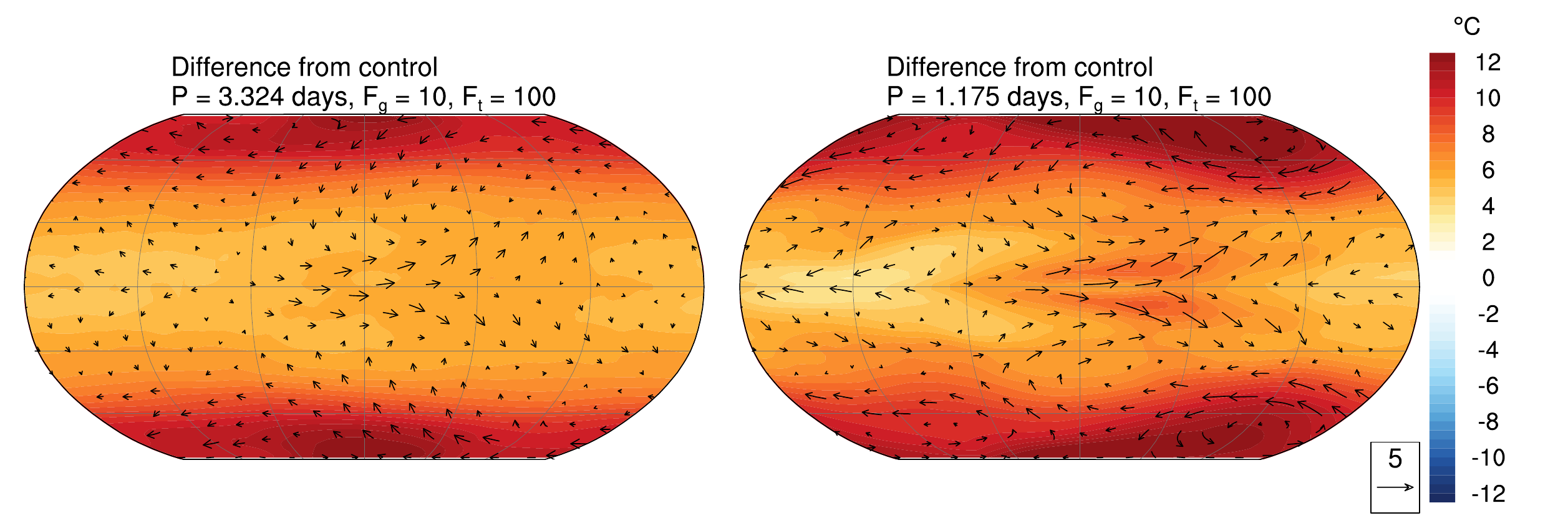}
\caption{Difference from control cases of the time average of surface temperature (shading) and horizontal wind (vectors) for the $P=3.324$\,d (left panel) and $P=1.175$\,d (right panel) experiments with $F_{\rm g} = 10\,{\rm W\,m}^{-2}$ and $F_{\rm t} = 100\,{\rm W\,m}^{-2}$. The lengths of the vectors are proportional to the local wind speeds, and a reference vector with a length of 5\,m\,s$^{-1}$ is shown on the last panel.}
\label{fig:surftemp_longshort}
\end{figure*}
%**********************************************

Polar amplification is commonly attributed to the warming that results from ice-albedo feedback in polar regions,
which accelerates the loss of ice, thereby reducing albedo and continuing to accelerate the rate of warming \citep{Polyakov2002,Holland2003}.
However, polar amplification is also present in idealized models that lack sea ice feedback entirely \citep{Alexeev2005,langen2007polar,alexeev2013polar},
which suggests that atmospheric heat transport alone can provide an explanatory mechanism. \citet{Alexeev2005} demonstrated that 
polar amplification should occur when a GCM is forced with an additional source of uniform surface warming, which in principle could be casued by  
both infrared and visible sources. 
\citet{alexeev2013polar} argued that both ice-albedo feedback
and meridional energy transport contribute to polar amplification, with the effects of energy transport being masked when ice-albedo feedback is present.

Our exomoon simulations further illustrate the capability of GCMs to show polar amplification in the absence of ice-albedo feedback.
Previous idealized GCMs have shown polar amplification from uniform thermal heating sources, and our results demonstrate that similar polar amplification
can be obtained from {the} non-uniform thermal heating {of planetary illumination.}
Polar amplification on ice-free planets occurs as a response to enhancement of the meridional circulation on a warmer planet \citep{Lu2007}, which leads to increased 
poleward transport of energy and moisture \citep{Alexeev2005}. 

{We calculate the polar amplification, ${\Delta}T_{\rm pole}$ as the difference in the mean temperature at the north pole
between each experiment and the corresponding control case. The values of ${\Delta}T_{\rm pole}$ are shown in Table~\ref{tab:models}.
A constant uniform geothermal heating of $F_{\rm g} = 10$\,W\,m$^{-2}$ with no planetary illumination (cases 3 and 8) yields a 
polar amplification of about $4\,K$, with the outgoing longwave radiation also about $10$\,W\,m$^{-2}$ greater than the control case.
This indicates that geothermal heating is entirely absorbed and re-radiated by the lower, thicker layers of the atmosphere.
Conversely, a non-uniform planetary illumination of $F_{\rm t} = 10$\,W\,m$^{-2}$ with no geothermal heating (cases 2 and 7)
shows a smaller polar amplification of about $0.6\,K$. {Cases 2 and 7 show} 
an increase in outgoing longwave radiation of only about $3$\,W\,m$^{-2}$ compared
to the control; {however, this is expected because the non-uniform distribution of planetary heating 
($F_{\rm t}\left|\cos\lambda\right|$ when $90^\circ < \lambda < 270^\circ$)
results in a net warming of $F_{\rm t}/\pi \approx 3.2$\,W\,m$^{-2}.$
Even so, the total polar amplification of $0.6\,K$ in cases 2 and 7 is nearly seven times less than
the polar amplification of $4\,K$ in cases 3 and 8.}
This indicates that {some} planetary illumination is absorbed by the 
uppermost layers of the atmosphere, with {the remainder} of this energy contributing
to surface warming. However, these cases (2 and 7) still demonstrate that polar amplification can occur from a non-uniform upper-atmospheric
heating source. The experiments with equal geothermal and planetary heating of $F_{\rm g} = 10$\,W\,m$^{-2}$ and $F_{\rm t} = 10$\,W\,m$^{-2}$ 
(cases 4 and 9) show that the value of ${\Delta}T_{\rm pole}$ is equal to the sum of the polar amplification when each heating source is considered in isolation.
Likewise, the value of $F_{\rm OLR}$ for case 4 (and 9) equals the sum of the outgoing longwave radiation terms from cases 2 and 3 (7 and 8).
Polar amplification continues to increase when $F_{\rm t} = 100$\,W\,m$^{-2}$ and greater (cases 5, 10-12), with corresponding 
increases in $F_{\rm OLR}$ that indicate a further penetration depth for incoming planetary illumination. 
These results emphasize that polar amplification in GCMs that lack ice albedo feedback can still occur 
with both uniform surface warming and non-uniform stratospheric warming.}

{The expansion of the meridional overturining (i.e., Hadley) circulation is shown in Fig. \ref{fig:MMC-long}
for the $P = 3.324$\,d experiments. The top row of Fig. \ref{fig:MMC-long} shows the control case, while the bottom 
row shows the experiment with $F_{\rm g} = 10$\,W\,m$^{-2}$ and $F_{\rm t} = 100$\,W\,m$^{-2}$. 
The left column of Fig. \ref{fig:MMC-long} shows the global average, while the middle and right columns separate the
atmosphere into the hemispheres east and west, respectively, of the subplanetary point. 
The purpose of this decomposition is to show that the atmosphere responds to a fixed heating source by altering both the 
direction and width of the Hadley circulation in each hemisphere relative. This prediction originates from the shallow 
water model of \citet{Geisler1981}, which demonstrated that the Hadley circulation should change directions on either side 
of a fixed heating source. Although the Hadley circulation appears to weaken and maintain its width when planetary and geothermal 
heating is applied (left column), the hemispheric decomposition shows that the Hadley circulation in eastern hemisphere reaches 
fully to the poles while the western hemisphere shows a Hadley circulation with the opposite direction. The zonal wind pattern
in these atmospheres remains relatively consistent between the two cases, with prominent upper-level jets associated with the 
descending branch of the global Hadley cell that appear identical in the hemispheric decomposition.}

%**********************************************
\begin{figure*}
\centering
\includegraphics[width=1.\linewidth]{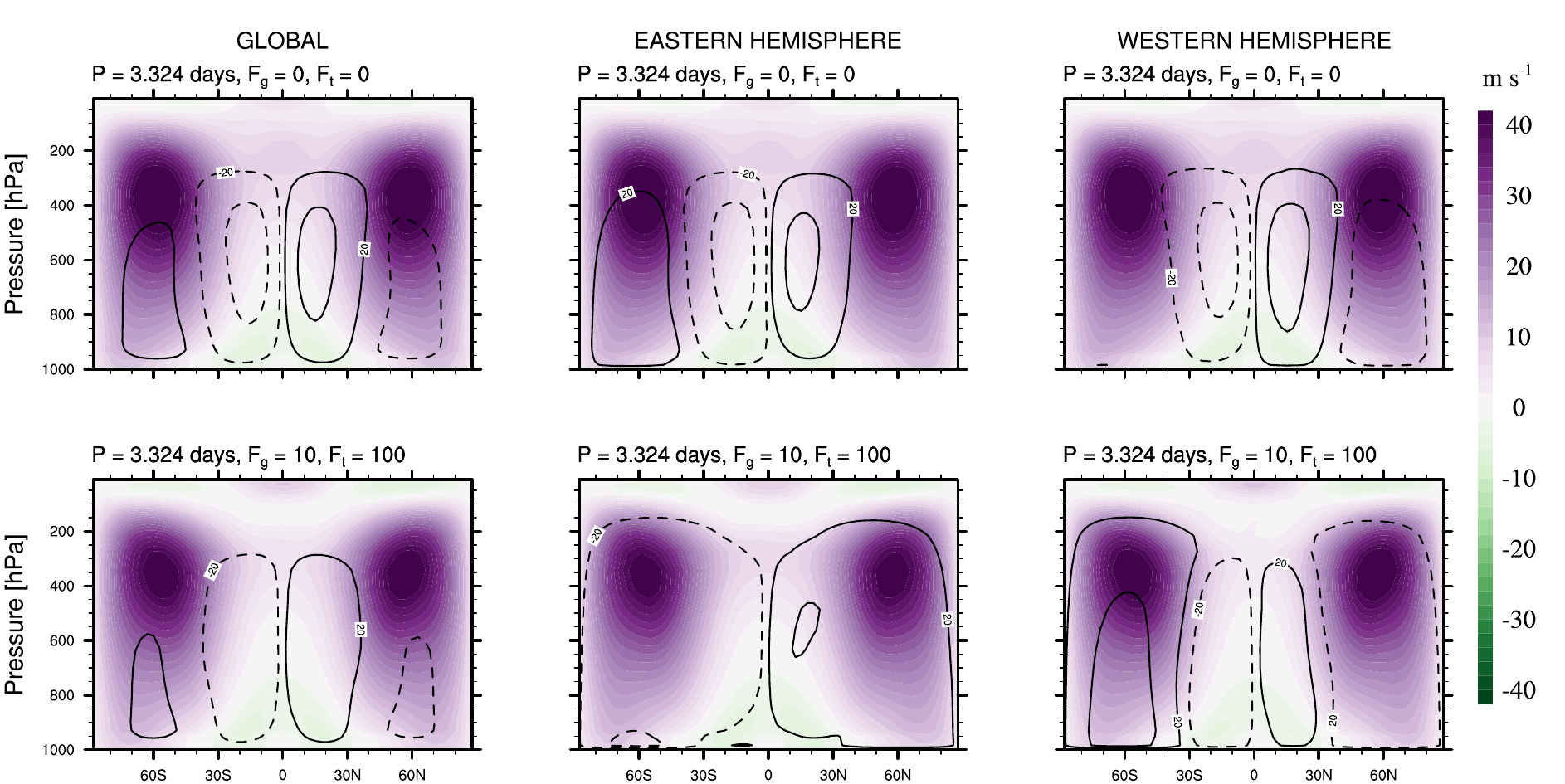}
\caption{The mean meridional circulation (line contours) and zonal mean zonal wind
(shading) are averaged across the entire planet (first column), eastern
hemisphere (second column), and western hemisphere (third column)
from the subplanetary point for the 
$P = 3.324$\,d experiments with $F_{\rm g} = F_{\rm t} = 0$\,W\,m$^{-2}$ (top row)
and $F_{\rm g} = 10$\,W\,m$^{-2}$ and $F_{\rm t} = 100$\,W\,m$^{-2}$ (bottom row). Contours are drawn
at an interval of $\pm\{20,100,300\}\times10^{9}\text{ kg s}^{-1}$.
Solid contours indicate positive (northward) circulation, and dashed contours
indicate negative (southward) circulation.}
\label{fig:MMC-long}
\end{figure*}
%**********************************************

{We also demonstrate this Hadley cell expansion or the $P = 1.175$\,d experiments in Fig. \ref{fig:MMC-short},
which shows the control experiment along with three other cases of increasing planetary illumination 
and geothermal heating. Fig. \ref{fig:MMC-short} shows that the global average Hadley circulation tends 
to diminish as planetary illumination increases, but the hemispheric decomposition shows that 
the Hadley circulation is actually expanding poleward. The two hemispheres show circulations with opposite 
direction and approximately the same strength. Geothermal heating does not significantly alter the circulation strength,
although it does change the morphology of the ascending branch of the Hadley circulation. Two upper-level midlatitude jets are 
present when $F_{\rm t} \le 100$\,W\,m$^{-2}$, while a third equatorial jet emerges when $F_{\rm t} = 500$\,W\,m$^{-2}$. 
Strong geothermal heating of $F_{\rm g} = 100$\,W\,m$^{-2}$ tends to sharpen the equatorial jet and raise the altitude 
of the midlatitude jets.}

%**********************************************
\begin{figure*}
\centering
\includegraphics[width=1.\linewidth]{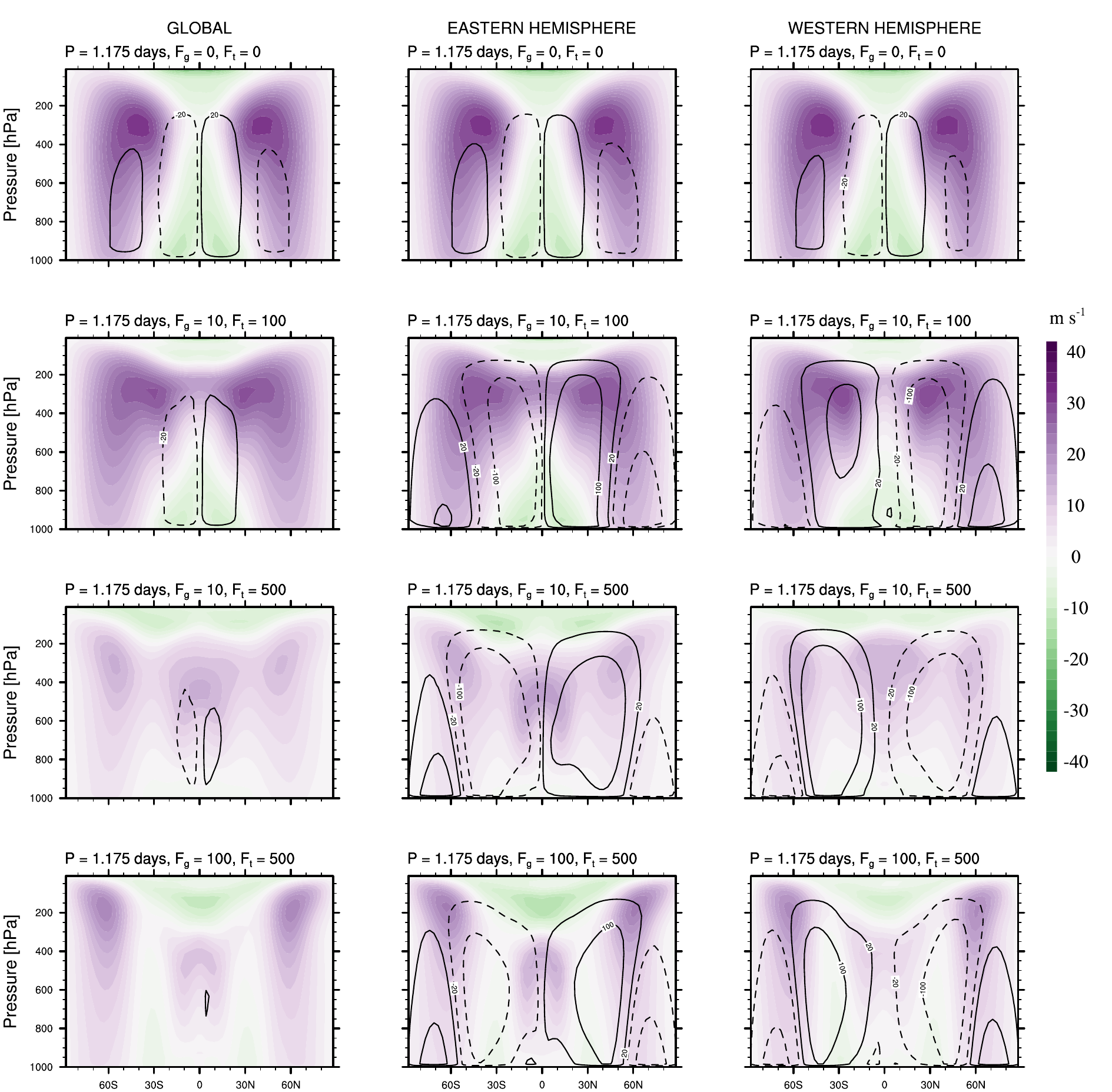}
\caption{The mean meridional circulation (line contours) and zonal mean zonal wind
(shading) are averaged across the entire planet (first column), eastern
hemisphere (second column), and western hemisphere (third column)
from the subplanetary point for the 
$P = 1.175$\,d experiments with $F_{\rm g} = F_{\rm t} = 0$\,W\,m$^{-2}$ (first row),
$F_{\rm g} = 10$\,W\,m$^{-2}$ and $F_{\rm t} = 100$\,W\,m$^{-2}$ (second row),
$F_{\rm g} = 10$\,W\,m$^{-2}$ and $F_{\rm t} = 500$\,W\,m$^{-2}$ (third row),
and $F_{\rm g} = 100$\,W\,m$^{-2}$ and $F_{\rm t} = 500$\,W\,m$^{-2}$ (last row). 
Contours are drawn at an interval of $\pm\{20,100,3000\}\times10^{9}\text{ kg s}^{-1}$.
Solid contours indicate positive (northward) circulation, and dashed contours
indicate negative (southward) circulation.}
\label{fig:MMC-short}
\end{figure*}
%**********************************************

We {further} demonstrate this behavior in our results by examining the vertically integrated 
flux of moist static energy as a function of latitude, following \citet{Frierson2007b} and \citet{Kaspi2015}. 
Moist static energy, $m$, represents the combination of dry 
static energy and latent energy as 
\begin{equation}
m = c_pT + \Phi + L_vq, 
\label{eq:MSE}\end{equation}
where $c_p$ is the specific heat capacity of air, $\Phi$ is geopotential height, 
$L_v$ is the enthalpy of vaporization, and $q$ is specific humidity. 
We decompose moist static energy
$m$ into a sum of time mean $\overline{m}$ and eddy $m'$ contributions
as $m=\overline{m}+m'$.
This allows us to write the meridional moist
static energy flux as 
\begin{equation}
\overline{vm}=\bar{v}\bar{m}+\overline{v'm'},
\label{eq:vmbar}\end{equation}
where $\overline{v}\overline{m}$ represents meridional mean energy
transport and $\overline{v'm'}$ represents meridional eddy energy
transport.
The vertically integrated flux of $m$ is defined as 
\begin{equation}
\overline{M} = 2\pi a\cos\phi\int_{0}^{p_{s}}\frac{\overline{vm}}{g}dp, 
\label{eq:MSEflux}\end{equation}
where $a$ is planetary radius, $\phi$ is latitude, and the overbar denotes a zonal 
and time mean. Equation~(\ref{eq:MSEflux}) gives the total value of $\overline{M}$ from all 
dynamical contributions. We can likewise separate the mean and eddy contributions to $\overline{M}$ 
by replacing the meridional static energy flux $\overline{vm}$ in Eq. (\ref{eq:MSEflux}) with
$\bar{v}\bar{m}$ or $\overline{v'm'}$, respectively.

We present the total, mean, and eddy fluxes of $\overline{M}$ in Figure~\ref{fig:MSE} to show the 
difference between our control cases and our experiments with $F_{\rm g} = 10$\,W\,m$^{-2}$ and $F_{\rm t} = 100$\,W\,m$^{-2}$.
Both panels show a poleward increase in $\overline{M}$ when geothermal and planetary heating are added, 
with all of this contribution due to increases in the mean component of $\overline{M}$ at latitudes $\phi > 50^{\circ}$. By contrast, the eddy component of $\overline{M}$
decreases in the range $30^{\circ} < \phi < 50^{\circ}$ when heating is induced. Polar amplification, and an associated 
increase in moisture, occurs as a result of intensification of the mean poleward transport of static and latent energy fluxes
from both surface geothermal and top-of-atmosphere planetary heating.

%**********************************************
\begin{figure*}
\centering
\includegraphics[width=1.\linewidth]{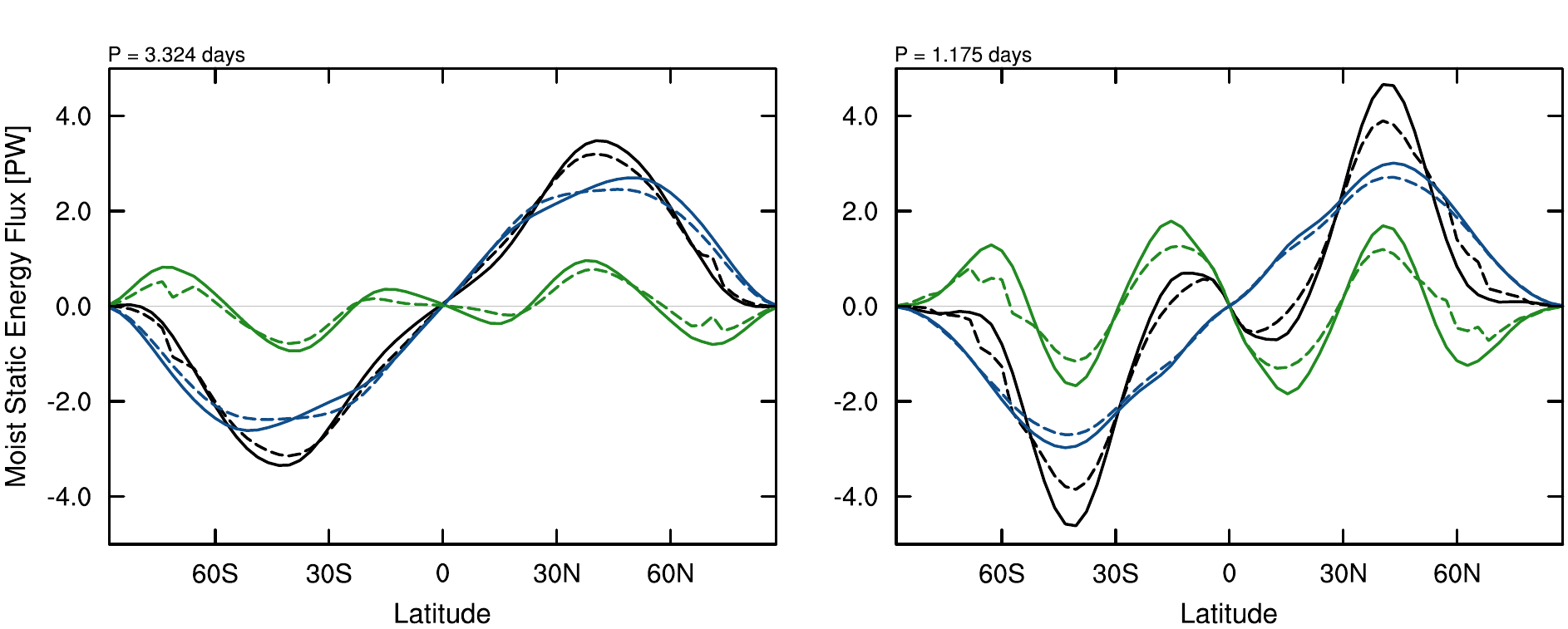}
\caption{Vertically integrated moist static energy flux ($\overline{M}$) for the $P = 3.324$\,d (left panel) and $P = 1.175$\,d (right panel) experiments. The total $\overline{M}$ (black curves), mean contribution to $\overline{M}$ (green curves), and eddy contribution to $\overline{M}$ (blue curves) are shown for control cases with $F_{\rm g} = F_{\rm t} = 0$ (solid) and experiments with $F_{\rm g} = 10\,{\rm W\,m}^{-2}$ and $F_{\rm t} = 100\,{\rm W\,m}^{-2}$ (dashed).}
\label{fig:MSE}
\end{figure*}
%**********************************************

The moist static energy flux also provides an explanation for the equatorial band of warming in our difference plots shown in Figure~\ref{fig:surftemp_longshort}, particularly
in the rapid rotating case. Figure~\ref{fig:MSE} also shows a decrease in $\overline{M}$ at tropical latitudes in the range $0^{\circ} < \phi < 15^{\circ}$, which leads 
to warming and an accumulation of moisture beneath the subplanetary point. This point corresponds to a maximum in the rising motion of the mean meridional
circulation (not shown), as a result of planetary illumination. Although our GCM does not include cloud processes, convective processes at the subplanetary point 
should lead to cloud formation, which could contribute to an expansion of the inner habitable region around the host planet \citep{2014ApJ...787L...2Y}.
In general, the circulation patterns of climates with a fixed source of heating are not 
easily characterized by longitudinally-averaged mean meridional circulation functions, due to hemispheric reversals in the direction  
of these circulation patterns \citep{2015MNRAS.446..428H}. Nevertheless, we can still expect strong rising motion beneath the subplanetary point, with 
both zonal and meridional transport toward the opposing hemisphere.

%**********************************************
\begin{figure*}
\centering
\includegraphics[width=1.\linewidth]{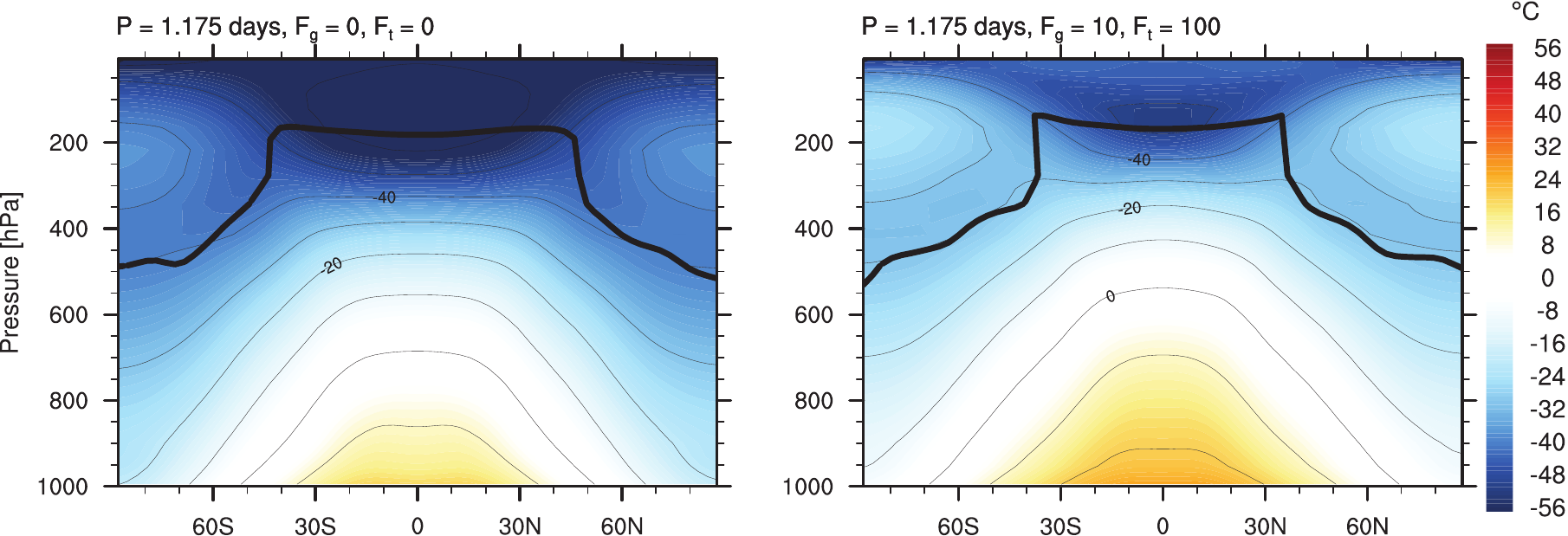}
\caption{Time average of mean zonal temperature (shading and contours) for the $P = 1.175$\,d experiments with $F_{\rm g} = F_{\rm t} = 0$\,W\,m$^{-2}$ (left panel) and $F_{\rm g} = 10$\,W\,m$^{-2}$ and $F_{\rm t} = 100$\,W\,m$^{-2}$ (right panel). The dark curve shows the height of the tropopause. (The color scale is chosen to match previous figures, and contours are drawn every $10\,^{\circ}$C.)}
\label{fig:temptrop}
\end{figure*}
%**********************************************

\subsection{Vertical structure of the atmosphere}

The redistribution of energy from planetary illumination causes warming of both the surface as well as the polar stratosphere. 
Figure \ref{fig:temptrop} shows the vertical structure of mean zonal temperature for the short-peroid cases with $F_{\rm g} = F_{\rm t} = 0$\,W\,m$^{-2}$ (left panel) and
$F_{\rm g} = 10$\,W\,m$^{-2}$ and $F_{\rm t} = 100$\,W\,m$^{-2}$ (right panel). The height of the tropopause is also shown as a dark curve in both panels of Figure \ref{fig:temptrop}, 
which follows the World Meteorological Organization definition of the tropopause as the altitude at which the lapse rate equals $-2$\,K\,km$^{-1}$. For the control case (left panel), 
the tropical tropopause extends up to about $100$\,hPa due to convective heating by both moist and dry processes \citep{Haqq-Misra2011}. The height of the 
extratropical tropopause is determined by the balance between warming from latent and sensible heating in the troposphere below with warming in the stratosphere 
from the poleward transport of the Brewer-Dobson circulation \citep{Haqq-Misra2011}. 

When planetary and geothermal heating are included (Figure \ref{fig:temptrop}, right panel), the boundaries of the tropical tropopause sharpen and act to widen the extratropical 
zone while narrowing the tropics. Increased warming beneath the subplanetary point drives stronger convection, which causes the tropopause to extend higher. 
This increase in convective heating also increases the poleward flux of moist static energy, which causes the extratropical tropopause to steepen. Warming in the stratosphere 
occurs primarily in the polar regions, which is driven by poleward energy transport processes such as the Brewer-Dobson circulation (not shown). This stratospheric warming  
competes with tropospheric warming in the tropics, which results in the polar height of the tropopause remaining relatively constant between the two cases shown.
One interpretation of this behavior is that the expanded extratropics are analogous to an increase in efficiency of the moon's `radiator fins,' which provide 
a means of transporting and dissipating energy from the warming tropics \citep{pierrehumbert1995}.
Planetary illumination thus serves to sharpen the distinction between tropical and extratropical climate zones by redistributing energy poleward both along the surface 
and aloft in the stratosphere.

The effect of planetary illumination on the atmospheric structure is also evident from examining vertical temperature profiles along the equator (Fig. \ref{fig:tempprofiles}{, left panel)}. 
For all profiles, the long-period cases show a colder stratosphere but a warmer troposphere than the short-period cases with the same planetary and geothermal heating. 
These differences correspond to the enhanced meridional transport of energy and moisture in the
long-period cases, which also drives stronger and narrower jets in the short-period cases \citep{Williams1982}.
The effect of planetary illumination causes a stratospheric temperature inversion, analogous to the ozone-driven stratospheric inversion on Earth today. 
The warmest cases with $F_{\rm t} = 500$\,W\,m$^{-2}$ show a profile with increasing slope that begins approaching an isothermal atmosphere from the additional 
planetary and geothermal heating.

%**********************************************
\begin{figure*}
\centering
\includegraphics[width=1.\linewidth]{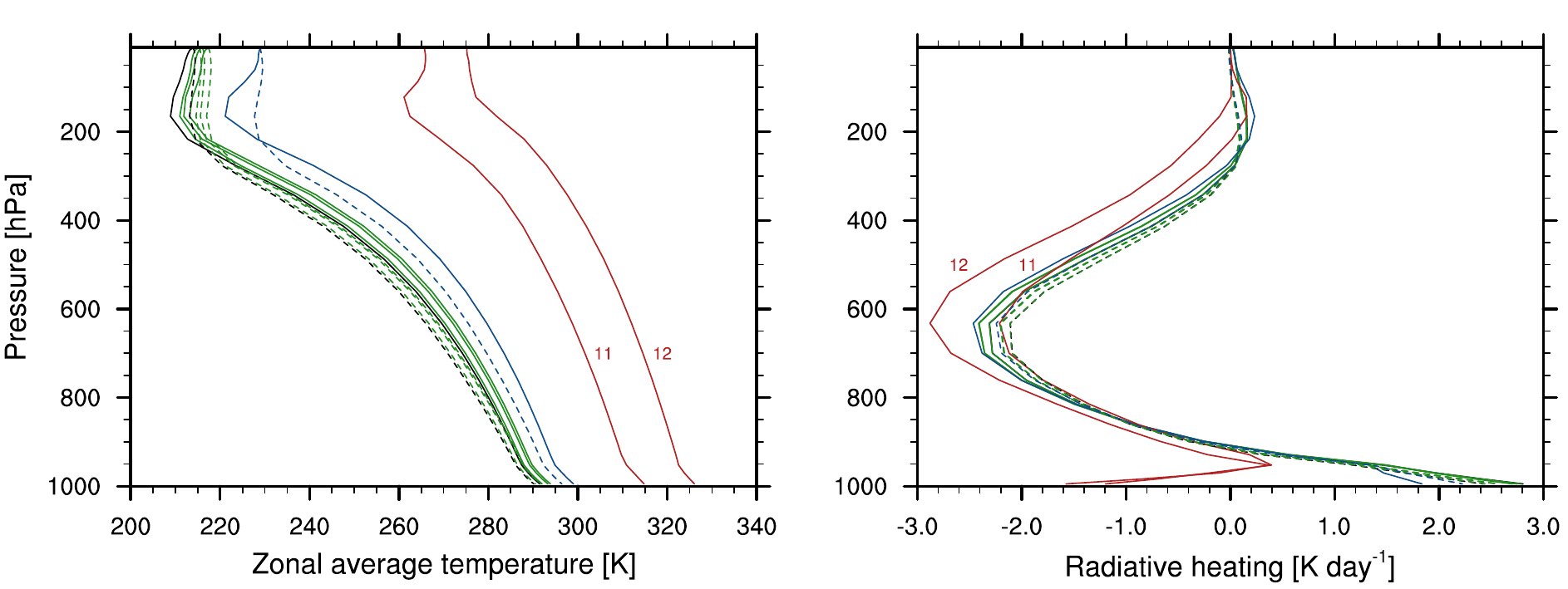}
\caption{Vertical profiles of the zonal mean temperature {(left) and radiative heating (right)} at the equator for all simulations listed in Table~\ref{tab:models}.
Solid curves indicate short-period cases ($1.175$\,d) and dashed curves indicate long-period cases ($3.324$\,d).
Black curves show control experiments, while green indicates all simulations with $F_{\rm t} \le 10$\,W\,m$^{-2}$.
Blue curves show simulations with $F_{\rm g} = 10$\,W\,m$^{-2}$ and $F_{\rm t} = 100$\,W\,m$^{-2}$, while red curves
show cases with $F_{\rm t} = 500$\,W\,m$^{-2}$. Case numbers as per Table~\ref{tab:models} are given for some of the lines.}
\label{fig:tempprofiles}
\end{figure*}
%**********************************************

{Some planetary illumination is absorbed in the uppermost layers of the model atmosphere, while the rest contributes to surface warming.
Fig. \ref{fig:tempprofiles} ({right panel)} shows vertical profiles of the temperature tendency due to radiation (or radiative heating) along the equator.
This reflects the direct change in temperature due to radiation alone, neglecting physical processes such as convection, boundary layer diffusion, and dynamical heating. 
The green curves with $F_{\rm t} = 10$\,W\,m$^{-2}$ show modest warming in the upper atmosphere of about $0.1$\,K\,day$^{-1}$, with strong cooling in the middle troposphere
and surface heating of about $2.8$\,K\,day$^{-1}$. The blue curves show increased upper atmosphere warming to $\sim0.2$\,K\,day$^{-1}$ with stronger cooling
in the middle troposphere. The blue curves also show a lower radiative heating rate at the surface, even though these cases show a higher surface temperature.
The decrease in direct surface radiative heating is accounted for by increased diffusion of the boundary layer, which in turn leads to increased
surface warming. The structure of the boundary layer is also evident from the left panel of Fig. \ref{fig:tempprofiles} as a change in lapse rate
near $\sim950$\,hPa. The middle troposphere is characterized by strong convection, which acts to restore the radiative cooling.}

{The more extreme cases with $F_{\rm t} = 500$\,W\,m$^{-2}$ also continue this trend, with radiative cooling of $\sim1.2$\,K\,day$^{-1}$ 
indicating a transfer of energy to the vertical diffusion of the atmosphere's boundary layer---which thereby leads to surface warming. Note
also that case 12 shows reduced upper-atmosphere absorption due to the large geothermal flux of $F_{\rm g} = 100$\,W\,m$^{-2}$ along with
stronger cooling (and thus convection) in the middle troposphere. The gray-gas radiative absorber in this GCM assumes a specified
vertical profile as a function of optical depth, which represents a greenhouse effect in the troposphere and stratosphere. The upper 
atmosphere absorption of planetary illumination on an actual exomoon would depend upon the atmospheric composition and pressure, among 
other factors, which could be explored with other GCMs that use band-dependent radiative transfer. Comparison of several different GCMs in
a similar exomoon configuration would provide more robust constraints on the expected heating profile from planetary illumination.}

\section{Discussion}

In general, these simulations illustrate that the potential habitability of an exomoon depends upon 
the thermal energy emitted by its host planet. 
We find stable climate states for both slow and rapid rotators with $F_{\rm g} \le 10$\,W\,m$^{-2}$ and $F_{\rm t} \le 100$\,W\,m$^{-2}$, 
which indicates that both geothermal heating and planetary illumination could provide an additional source of warming 
for an exomoon. However, strong thermal illumination by the host planet ($F_{\rm t} = 500$\,W\,m$^{-2}$ in our experiments) 
would likely lead to an onset of a runaway greenhouse and the loss of all standing water. 

On the one hand, the habitability of some exomoon systems may therefore be precluded based upon the presence of a luminous host planet, although planetary illumination itself  does not necessarily limit an exomoon's habitability \citep{2016A&A...588A..34H}.
On the other hand, we expect polar amplification of warming in all cases, which may suggest that exomoons in orbit around a luminous host planet may be less likely to develop polar ice caps. This tendency for an exomoon to have warmer poles due to planetary illumination suggests that such bodies may have a greater fractional habitable area than Earth today \citep{2008ApJ...681.1609S}, potentially improving the prospects of an origin of life \citep{2014AsBio..14...50H}. 
This prediction of polar warming on exomoons could eventually translate into observables from the circumstellar phase curves of an exomoon, if the planet's contribution to the combined planet-moon phase curve can be filtered out \citep{2012ApJ...757...80C,2017MNRAS.470..416F}.

To a lesser extent, the rotational period of the exomoon also contributes to differences in surface habitability. The short-period cases tend 
to show a greater amount of warming along the equator, near the subplanetary point (Figure~\ref{fig:surftemp_longshort}, right panel), 
which could serve as an additional source of energy to maintain regional habitable conditions. Dynamical 
changes in atmospheric jet structure that result from differences in rotation rate also contribute to changes in surface wind patterns,
as well as the general circulation, which will likely correspond to significant contrasts in resulting cloud patterns. 
{The large-scale circulation also shows an opposite directional sense in the hemispheres east and west of the subplanetary illumination 
point, which could also impact the probable location of clouds.}	

{The hemispheric differences in the Hadley circulation (Figs. \ref{fig:MMC-long} and \ref{fig:MMC-short}) show similarities to simulations of terrestrial planets in synchronous
rotation around low mass stars, where the host star is fixed upon a substellar point on the planet. 
\citet{Haqq-Misra2011} used the idealized FMS GCM to demonstrate that the Hadley circulation shows direction in the opposite direction when comparing the hemispheres 
east and west of the substellar point for planets with 1\,d and 230\,d rotation periods. 
\citet{Haqq-Misra2018} also find the same hemispheric circulation patterns in an analysis the Community Earth System Model (CESM), which includes band-dependent radiation, 
cloud processes, and other physical processes. Synchronously rotating planets drive such a circulation when their stellar energy source is fixed to a single 
location; however, our exomoon calculations demonstrate that such a circulation can also be obtained when planetary illumination is fixed,
even if the moon otherwise experiences variations in incoming starlight.}

These idealized calculations provide a qualitative description of surface temperature and winds on an exomoon, but the application of 
more sophisticated GCMs will help to identify particular threshold where a runaway greenhouse and other water loss processes occur.
Non-gray radiative transfer will allow for particular atmospheric compositions to be examined, such as a mixture of nitrogen, carbon dioxide, and 
water vapor that is characteristic of Earth-like atmospheres. Implementation of a cloud scheme into the GCM will also provide 
important insights into habitability, as clouds could help to delay the onset of a runaway greenhouse state \citep{2014ApJ...787L...2Y}.

{In terms of the odds of an actual detection of the climatic effects described in this paper, this could in principle be possible if the moon's electromagnetic spectrum (either reflection or emission) could be separated from that of the planet. This might be possible in very fortunate cases where a large moon is transiting its luminous giant planet \citep{2014ApJ...796L...1H,2016A&A...588A..34H} or where both the planet and its moon transit their common low-mass host star \citep{2010ApJ...712L.125K}. Alternatively, if the moon is subject to extreme tidal heating, it could even outshine its host planet in the infrared and therefore directly present its emission spectrum \citep{2013ApJ...769...98P} while the planet would still dominate the visible part of the spectrum, where it reflects much more light than the moon. The technological requirements, however, will go beyond the ones offered by the James Webb Space Telescope \citep{2010ApJ...712L.125K} and might not be accessible within the next decade.}

\section{Conclusions}
We present the first GCM simulations of the atmospheres of moons with potentially Earth-like surface conditions. Our simulations illustrate the effects of tidal heating and of planetary illumination on the atmospheres of large exomoons that could be abundant around super-Jovian planets in the stellar habitable zones. 

Most of the energy from planetary thermal heating and geothermal heating 
is transported toward the poles as a result of enhanced meridional transport 
of moisture and energy. This suggests that polar ice melt may be less prevalent 
on exomoons that are in synchronous rotation with their host planet. In general, these calculations further illustrate that 
{the poleward expansion of the Hadley circulation enhances meridional energy transport and can lead to polar amplification of warming, 
even in the absence of ice albedo feedback.}

This polar heat transport could increase the fraction of the surface that allows the presence of liquid surface water by compensating for the lower stellar flux per area received at the poles of a moon. In other words, illumination from the planet might be beneficial for the development of life on exomoons. Future observations that are able to distinguish exomoons from their host planet may be able to detect the absence of polar ice caps due to polar amplification of planetary illumination, such as analysis of photometric phase-curves.

\section*{Acknowledgments}
The authors thank Ravi Kopparapu for helpful feedback on a previous version of the manuscript. J.H. acknowledges funding from the NASA Astrobiology Institute's Virtual Planetary Laboratory under awards NNX11AC95G and NNA13AA93A, as well as the NASA Habitable Worlds program under award NNX16AB61G. R.H. has been supported by the German space agency (Deutsches Zentrum f\"ur Luft- und Raumfahrt) under PLATO Data Center grant 50OO1501, by the Origins Institute at McMaster University, and by the Canadian Astrobiology Program, a Collaborative Research and Training Experience Program funded by the Natural Sciences and Engineering Research Council of Canada (NSERC). Any opinions, findings, and conclusions or recommendations expressed in this material are those of the authors and do not necessarily reflect the views of NASA or NSERC.

%%%%%%%%%%%%%%%%%%%%%%%%%%%%%%%%%%%%%%%%%%%%%%%%%%



\bibliographystyle{mnras}
\bibliography{exomoon_ms} % if your bibtex file is called example.bib

%%%%%%%%%%%%%%%%%%%%%%%%%%%%%%%%%%%%%%%%%%%%%%%%%%

% Don't change these lines
\bsp	% typesetting comment
\label{lastpage}
\end{document}